# Designing Tools for Semi-Automated Detection of Machine Learning Biases: An Interview Study

**Po-Ming Law**
Georgia Institute of Technology
Atlanta, GA, USA
pmlaw@gatech.edu

**Sana Malik**
Adobe Research
San Jose, CA, USA
sana.malik@adobe.com

**Fan Du**
Adobe Research
San Jose, CA, USA
fdu@adobe.com

**Moumita Sinha**
Adobe Research
San Jose, CA, USA
mousinha@adobe.com

## ABSTRACT

Machine learning models often make predictions that bias against certain subgroups of input data. When undetected, machine learning biases can constitute significant financial and ethical implications. Semi-automated tools that involve humans in the loop could facilitate bias detection. Yet, little is known about the considerations involved in their design. In this paper, we report on an interview study with 11 machine learning practitioners for investigating the needs surrounding semi-automated bias detection tools. Based on the findings, we highlight four considerations in designing to guide system designers who aim to create future tools for bias detection.

**INTRODUCTION**

Machine learning (ML) pervades various aspects of everyday life. From powering advanced functionality in software applications to facilitating decision making in societal problems, ML models have been improving user experience and economic efficiency. However, these models often exhibit biases against some population subgroups. For instance, facial recognition software often performs poorly on certain skin types [8] or credit scoring models might reject loans much more frequently for certain racial groups [15].

Machine learning biases can have serious financial and ethical implications. In behavioral advertising, for example, ML models serve personalized advertisements by inferring user interests from behavioral data [6]. Poor predictive accuracy for certain marketing segments implies displaying less relevant advertisements, missing the opportunity to foster purchase intention. In criminal justice, models have been trained to help judges decide whether to detain defendants by predicting the likelihood of recidivism [5]. Predicting certain ethnic groups to be more likely to recommit crimes based on historical data defy our moral intuition.

To avoid consequences, ideally, ML practitioners should check their models for biases and mitigate these biases before deploying the models. In reality, it is difficult to detect ML biases in advance, and practitioners often stumble upon issues only after the deployment. The challenge lies in a large and unwieldy space of possible biases, making it onerous to ascertain the non-existence of significant biases. Even with thorough considerations, critical issues often slip through the cracks [12]. The difficulty in foreseeing ML biases has prompted scholars to call for more proactive model auditing processes [18].

Semi-automated tools that combine the strengths of humans and algorithms offer opportunities to support proactive bias detection [18]. These tools could mine biases from ML models and present them to practitioners for review. In this paper, we report on an interview study with 11 modelers (i.e. practitioners who train ML models) for investigating their needs for semi-automated bias detection tools. The contributions of this paper are two-fold:

- Findings from an interview study that aims to investigate the needs for semi-automated tools that involve humans in the detection of ML biases.
- A discussion of four considerations in designing semi-automated bias detection tools for practitioners who train ML models.

**BACKGROUND AND RELATED WORK**

While the ML community has proposed various types of ML biases, our interviews focus on subgroup biases: ML models can make predictions in ways that bias against certain subgroups of input data. Subgroup biases are common in ML systems. A well-known example is manifested in facial recognition systems. These systems, when employing algorithms trained on biased data (e.g., an image dataset in which people of color are underrepresented), have a low classification accuracy on certain ethnic groups [8].

ML bias research has centered on proposing statistical measures to quantify subgroup biases in ML classifiers [13, 16, 17, 23]. These measures are often adopted by algorithms for automatically detecting biases in ML models. While research in bias measures resides in the landscape of automated bias detection, our interviews concern the design of semi-automated tools that involve humans in bias detection and investigate the design considerations.

There has been development of semi-automated tools to help practitioners detect subgroup biases [2, 3, 4, 7, 9, 20, 21, 22]. However, these tools are often designed in isolation from users [18]. With an insufficient understanding of users, they may be divorced from user needs and expectations [19]. Our study aims to provide deeper insights into practitioners' needs for semi-automated bias detection tools. The findings could offer guidance to developers who want to design future tools and advance a conversation of how semi-automated bias detection tools could be designed.

The most relevant work is an interview study that dived into practitioners' needs for ML fairness [18]. Our study serves as an extension to the prior study by focusing on the needs for semi-automated bias detection tools. While the prior study suggested using semi-automated tools to facilitate fairness auditing, it did not delineate the considerations in designing these tools. Our study fills the gap by proposing four design considerations. By being more focused, our interviews revealed new insights. For instance, we found that modelers wanted tools that can automatically extract the causes of biases in the training data, which was not documented in the prior study.

## METHOD

To inquire into the needs for semi-automated bias detection tools, we conducted semi-structured interviews with practitioners who had experience in training ML models.

### Participants

We conducted ten semi-structured interviews with 11 modelers at a large technology company (one interview had two interviewees). As an inclusion criterion, interviewees are required to have experience in training ML models. Participants covered diverse product areas including personalized advertising (5 interviewees), computer visions (2), email marketing (2), recommender systems (1), and document viewer (1). They were recruited through snowball sampling. We reached out to personal contacts who worked on ML-driven products and asked them to refer colleagues.

### Interview

Each interview lasted 30—60 minutes and was conducted in person (6 interviews) or by teleconference (4 interviews). We recorded audio with interviewees' permission. During the interviews, we began by asking interviewees to describe a project they had recently worked on and might be relevant to "machine learning biases". We kept the definition of ML biases open to get a glimpse of interviewees' perceptions of biases. We then depicted examples of subgroup biases with the commonly-used adult dataset [1]. We emphasized that subgroup biases include a wide range of scenarios where a model makes predictions that bias against certain subgroups. As practitioners hold diverse views on what it means by ML biases [10, 11], the examples helped stay on topic. Next, we sought to understand the subgroup biases they encountered by asking them to elaborate on the previous project they described (if they believed that the project was related to subgroup biases) or to depict other relevant projects (if they thought that the previously described project was not relevant). We then asked about the current practice of bias detection and how semi-automated bias detection can improve the workflow.

**Analysis**

Audio from the interviews was manually transcribed by the research team. A researcher open-coded the transcripts for themes. Throughout the coding, the coder discussed the codes with the rest of the research team and refine the codes through multiple iterations.

**Limitations**

By recruiting within a single technology company, our study sample might not be representative of the whole population of modelers. For instance, the viewpoints of practitioners who develop other types of products are not represented. However, we strove to ensure that interviewees covered diverse product types. Despite the limited sample, our findings could still provide insights into the space of possible needs around semi-automated bias detection tools.

**RESULTS**

In this section, we summarize four themes that emerged from the interviews.

**Types of Machine Learning Biases**

The ML biases depicted by interviewees could be broadly categorized into pure performance issues (do not involve fairness concerns) and group unfairness (involve fairness concerns).

Pure Performance Issues. The majority of interviewees (7/11) described how their models underperformed on certain subgroups of input data. These performance issues were unrelated to social unfairness and discrimination that are central to the discourse on ML fairness. P6 trained models to classify customers into marketing segments and serve offers to customers based on the segments to which they belong. These marketing segments (e.g., professional users and hobbyists) did not involve sensitive demographic information (e.g., ethnicity), freeing the performance issues from concerns about social unfairness. P6 said that the consequence of the inaccurate segmentation was often modest: *"The marketing materials that you get might be less relevant."*

Group Unfairness. A few interviewees (4/11) recounted episodes about fairness issues in their models. A data scientist working on behavioral advertising (P11) expressed unease because the personalization algorithms his team developed could serve advertisements based on sensitive demographic information. Algorithms treating different demographic groups differently often raised concerns about social unfairness and violation of regulations. *"What gets more hairy is from their IP, we can determine geolocation [...] that's where I start worrying about biases [...] If you are a bank, you can't determine whether or not to show somebody a mortgage based on their geolocation [...] That's against the law"* (P11).

**Availability of Group Membership of Input Data**

Automatically detecting whether an ML model is biased against certain subgroups often involves segmenting a test set into subgroups and measures model performance on each to identify the subgroups on which the model underperforms [14]. A prerequisite for conducting this bias analysis is having the group membership of the input data. A few interviewees (5/11) reported issues in gathering the group membership. The group membership was often not directly available. Even if it was available, it might be too sensitive or unreliable to be used in bias analysis.

<u>Not Directly Available.</u> P1 developed GAN-based models that take videos of a person as an input to create volumetric 3D videos. P1 commented that metadata of the input videos (e.g., skin color of the person in the video) required manual effort to create, making it challenging to know if the models had inferior performance on certain subgroups of input data: *"For the volumetric video data, it's not like you have skin type data unless someone goes through it and annotates it [...] People have not taken the time to classify all the data into the different categories."*

<u>Available but Sensitive.</u> The data scientist who trained market segmentation models (P6) commented that while there were third parties that could provide demographic information of the customers, they decided not to use the information because of privacy concerns: *"It would be great to have that sensitive information there because I know that's very useful [...] It's kind of dangerous to have it just from a privacy perspective [...] We don't want anything to go wrong."*

<u>Available but Not Reliable.</u> P6 alluded to unreliability as another reason that they did not gather demographic information for bias analysis: *"We've tried out two different dealers of this [demographic] information and just from doing a lot of spot checks found that a lot of it was just untrustworthy. It was simply not correct. At that point, we decided we don't need to pay millions of dollars to get information that we don't even know is reliable."*

**Need for Automated Detection of Model Biases**

Most interviewees (10/11) cited that automated detection of group biases would be a useful addition to their ML workflow. P5 depicted how a semi-automated bias detection tool could help: *"I think it could be useful if you can provide every case [of biases] and then measure the amount of bias in each case, and sort and visualize them."* P7 commented on the importance of humans in the loop: *"At the end of the day, whether a pattern is a bias or not is a function of the domain. It would be very hard to determine without getting a human in the loop."* Interviewees discussed two reasons why they needed automated detection of model biases.

<u>Difficulty in Foreseeing Biases.</u> Some interviewees (3/11) explained the need for automated bias detection by alluding to the difficulty in foreseeing biases. P1 said, *"Biases are difficult for the ones that you don't know. If you know that there is a bias, then it is easy to solve [...] Typically, we just don't know."* When asked about how they knew about the biases in the first place, P5 and P10 commented that they often relied on hunch and domain knowledge to conduct spot checks. However, this approach was manual and tedious: *"So, just imagine we have a hunch that something is wrong, it is really hard just to go and pinpoint those [biases] one by one"* (P10).

<u>Understanding the Caveat of Models.</u> A few (3/11) commented that automating bias detection helped understand the caveats in applying their models. In P9's words, *"[With automated bias detection,] we have a little more context on why the model is performing as such [...] We know the caveats to the model before applying it to the real use cases."*

**Need for Automated Extraction of Causes in Training Data**

Besides the need for automated detection of model biases, many interviewees (9/11) highlighted the utility of automatically uncovering causes of biases in the training data. P9 and P11 explained that most biases are introduced by the training data: *"Most of the biases we've introduced to models is coming from the training data"* (P9), and *"If my model is showing me this strong bias, my first assumption is going to be that that bias comes from the data"* (P11). Interviewees believed that knowing about the causes of biases in the training data could help resolve the biases and communicate the biases to other stakeholders.

<u>Facilitating Resolution of Biases.</u> Some interviewees (6/11) said that learning about causes of biases in the training data could help debiasing the models. P5 explained, *"When we have to resolve a bias, providing biases in the outcome might not be enough to resolve this problem [...] If your tool can provide the reasons in the data set [training data], you may want to make it balance."*

<u>Aiding in Communication</u>. Two data scientists (P6 and P9) indicated that revealing causes of biases in the training data could facilitate communication. P6 noted that their stakeholders wanted to understand why their models underperformed on certain input data: *"What they [the users of the models] really want to know is the reasoning behind the model. The bias tool can help us tell that story [...] like `the data are the data' [...] We can only collect so much data [for an underperforming group]."*

**DESIGN CONSIDERATIONS**

Grounded in the four themes that emerged from the interviews, we propose four considerations in designing semi-automated tools to help ML modelers detect ML biases.

<u>Investigating the Nature of Biases.</u> In a typical design scenario, system designers strive to translate user needs into design requirements. When articulating user needs, it is crucial to understand the nature of the ML biases users care about as it could shape the design of bias detection tools. Our interview study revealed that ML biases ranged from group unfairness to pure performance issues. When fairness is a concern, designers could prioritize comprehensiveness and develop tools to help ensure the non-existence of biases that have serious consequences (e.g., harming a company's reputation when a model discriminates against ethnic minorities). When fairness is not a concern, designers could prioritize efficiency. For example, the tool could filter out biases on users' behalf so that fewer detected biases are presented for review.

Extracting Group Membership. Having the group membership of input data is crucial for determining if a model is biased against certain subgroups. For example, checking if a facial recognition model underperforms on people of color requires knowing the demographic information (e.g., ethnicity) of the input images. Yet, we learned that the group membership was often not readily available for bias analysis and gathering such information could be fraught with privacy concerns. When collecting group membership does not invoke privacy concerns, designers could investigate the subgroup biases modelers care about and develop techniques for automatically extracting relevant group membership from the input data. Using demographic information for model auditing, however, often raises privacy issues. Coarse-grained demographic information is often less sensitive and could be employed for bias analysis [18].

Managing an Unwieldy Space of Model Biases. Interviewees cited difficulty in foreseeing biases as a motivation for automating bias detection. Ensuring non-existence of ML biases entails checking a large number of subgroups and navigating through a huge space of possible biases. To manage the unwieldy space of model biases, flexible selection and intelligent filtering could be incorporated into the design of semi-automated bias detection tools. The tool could enable users to flexibly segment the test data into subgroups and choose different measures to evaluate model performance across the subgroups. It could also filter out statistically insignificant biases so that fewer biases are presented to modelers for inspection.

Knowing Motivations behind Understanding the Causes. Interviewees described different motivations for understanding the causes of biases in the training data. The design requirements of semi-automated bias detection tools hinge on such motivations. When understanding the causes is driven by a need to resolve the biases, system designers could develop tools that suggest possible ways to debias the ML models. When modelers need to communicate the causes of biases to other stakeholders, designers could consider methods for facilitating communication (e.g., easy ways to incorporate bias detection results into presentation tools). Hence, learning about why modelers want to understand the causes of biases is instrumental in needfinding.

**CONCLUSIONS**

This paper reports on an interview study that investigates the needs around semi-automated bias detection tools. We distilled four considerations involved in the design of these tools. We hope that our findings could serve as guidance to designers who aim to develop semi-automated bias detection tools that are useful in practice and facilitate a conversation of how bias detection tools that involve humans in the loop could be designed.